\documentclass[aps,prl,twocolumn,showpacs,epsfig,epsf,amssymb,showkeys,tightenlines]{revtex4-1}
\usepackage{graphicx}
\usepackage{amssymb}
\usepackage{amsmath}
\usepackage{subfigure}
\begin{document}
%\widetext

\title{ Heirarchical Self Assembly: Self Organized nano-structures in a nematically ordered matrix of self assembled polymeric chains.
}

\author{Shaikh Mubeena, Apratim Chatterji}
\email{apratim@iiserpune.ac.in}
\affiliation{
IISER-Pune, 900 NCL Innovation Park, Dr. Homi Bhaba Road,  Pune-411008, India.\\
}
\date{\today}
\begin{abstract}

We report many different nano-structures which are formed when model nano-particles of different sizes 
(diameter $\sigma_n$) are allowed to aggregate in a background matrix of semi-flexible self assembled polymeric
worm like micellar chains. 
%The nano-particles, modelled by interacting Lennard Jones spheres,
%self assemble in the  polymeric matrix with different values of self avoidance diameters of polymer chains. 
The different nano-structures are formed by the dynamical arrest of phase-separating mixtures of micellar monomers 
and nano-particles.  The different morphologies obtained are the result of an interplay of the available free volume, the elastic energy of deformation 
of polymers, the density (chemical potential) of the nano-particles in the polymer matrix and, of course, 
the ratio of the size of self assembling nano-particles and self avoidance diameter of polymeric chains.  We have used a hybrid  semi-grand canonical 
Monte Carlo simulation scheme to obtain the (non-equilibrium) phase diagram of the 
self-assembled nano-structures. We observe rod-like structures of nano-particles which get self assembled in the gaps 
between the nematically ordered chains as well as percolating gel-like network of conjoined nanotubes. 
We also find a totally unexpected interlocked crystalline phase 
of nano-particles and monomers, in which each crytal plane of nanoparticles is separated by planes of perfectly organized 
polymer chains. We identified the condition which leads to such interlocked crystal structure.      
We suggest experimental possibilities  of how the results presented in this paper could be used to obtain different nano-structures in the lab.
\end{abstract}
%\keywords{soft matter, flow, hydrodynamic simulations}

\maketitle

There is persistent interest in the controlled self assembly and growth of nano-structures of predefined morphology
and size starting from small constituent nano-particles (NP) \cite{Frenkel2011,Curk2014,Lopes2001,Whitesides1991,Lu2008,Zaccarelli2008,Araki2006,
Shchukin1999,Charleux2012,Glotzer2004,Grzelczak2010,Whitmer2013a,Pablo2012,Whitmer2013,Gupta2012,vermant2010,Samori2012,Katz2011,Starr2003,Sanat2009,
Spaeth2011,Starr2008,ethaya2012,Li2012,Horsch2005,Parviz2003,Ozin2009,Leontidis2003,Gharbi2014}.
A separate non-alligned interest of 
physicists is in the formation and properties of topological defects when large particles (large compared to the size 
and spacing between nematogens) are introduced in ordered liquid crystalline nematic and smectic phases 
\cite{Gharbi2014,Lubensky1999,Poulin1997,Weitz1999,Senyuk2013,Fukuda2009,McCoy2008,Musevic2008,Ravnik2009,Tkalec2009,Skarabot2008,Zhou2008,Stark2001}. 
Recents experiments have also explored 
the self organization of nano-particles in a background matrix of nematically ordered micellar phase, but constraints 
in the choice of size of NPs led to the following two scenarios: small NPs of $2-3$ nm diameter pervade the nematic chains themselves
and form a dispersion/solution, whereas, larger NPs of size $8-26$ nm get expelled by the elastic energy of ordered nematic phases
and aggregate at the grain boundaries between nematic domains \cite{Sharma2009,Sharma2010,Sharma2011}. The distance between 
adjacent nematic chains was $5.7$ nm in the experiments.   

Our present study spans across these two different research domains and we use computer simulations 
to investigate the heirarchical self assembly of NPs in the free volume between parallel chains of 
nematically ordered  worm-like micelles (WM). The micellar polymers are self-assembled themselves 
from monomeric beads and have a length and size distribution controlled by monomer density and 
temperature \cite{Cates1999,Berret2004,Berret2009}.  In a computer simulation, we are able to systematically vary the diameter, chemical 
potential of the spherical NPs as well as the excluded volume (EV) of  self-avoiding semiflexible 
chains in the matrix of the NPs self-organize. Thereby, we observe the effect of the above parameters, as
well as the elasticity of the background micellar matrix on the morphology and size of the NP-nano-structures. 
The nano-structured aggregates in turn increase the effective density of 
monomers constituting the background matrix and make them more nematically ordered with longer chains 
spanning the length of the system. 

At suitable densities and radii of NPs we get rod like aggregates of different aspect ratios shaped by
 the  geometry of the background matrix:  EV and elastic energy costs of accomodating the NPs amongst the 
semiflexible polymeric micelles  encourage the NPs to form aggregates even at moderate number densities. 
Since the background matrix is not only deformable but also prone to scission and recombination, 
neighbouring rod-like aggregates of NPs can also fuse at times forming porous percolating networks of extended 
tubular structures. In experimental realizations of our studies, these nano-structures could 
be stable due to van der Waals attraction even if the background micellar matrix is 
dissolved away by adding suitable ions in solution by reverse-micellization as in \cite{Hegmann2007,Wang2009}. 
To our surprise, we also get a perfectly crystalline phase spanning the simulation box where both NM and the WM 
forming monomers form alternate lines of NP and monomers forming a closed packed structures. 
In the following we describe the model of (a) self assembling WM chains (b) the NP and (c) the 
polymer-NP interaction. We then describe the different nano-structures obtained and summarize the conditions 
under which the different assembled structures are formed.

The unit of length for our simulations is the diameter $\sigma$ of monomers which self assemble to 
form worm-like {\em equilibrium} polymers at a temperature $k_B T$; we set $k_B T$ to be the unit of 
energy. The interaction potential between neighbouring monomers
is the sum of two body and three body terms, $V_2$ and $V_3$,  respectively, with $\epsilon =110 k_BT$ :
\begin{equation}
V_2 = \epsilon [ (\frac{\sigma}{r_2})^{12} - (\frac{\sigma}{r_2})^6 + \epsilon_1 e^{-a r_2/\sigma}]; 
\, \forall  r < r_c.
\label{eq1}
\end{equation}

\begin{equation}
%V_3(r_{12},r_{13}) = \epsilon_3 (1 - \frac{r_{12}}{\sigma_3})(1 - \frac{r_{13}}{\sigma_3}) sin^2(\theta); 
V_3 = \epsilon_3 (1 - \frac{r_{2}}{\sigma_3})^2(1 - \frac{r_{3}}{\sigma_3})^2 \sin^2(\theta); 
\, \forall r_{2},r_{3} < \sigma_3. 
\label{eq2}
\end{equation}

\begin{figure}
\includegraphics[width=0.5\columnwidth]{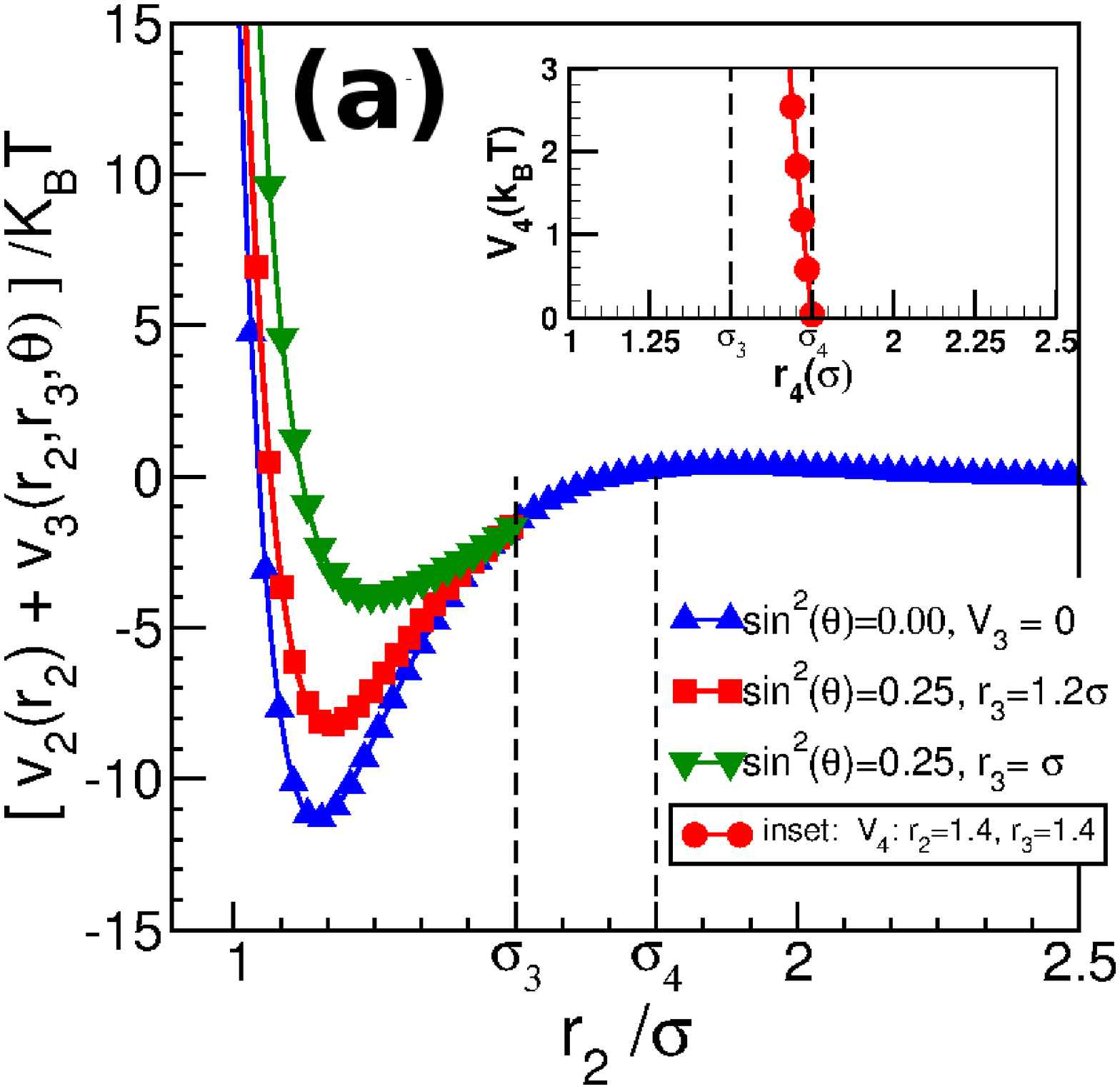}
\includegraphics[width=0.35\columnwidth]{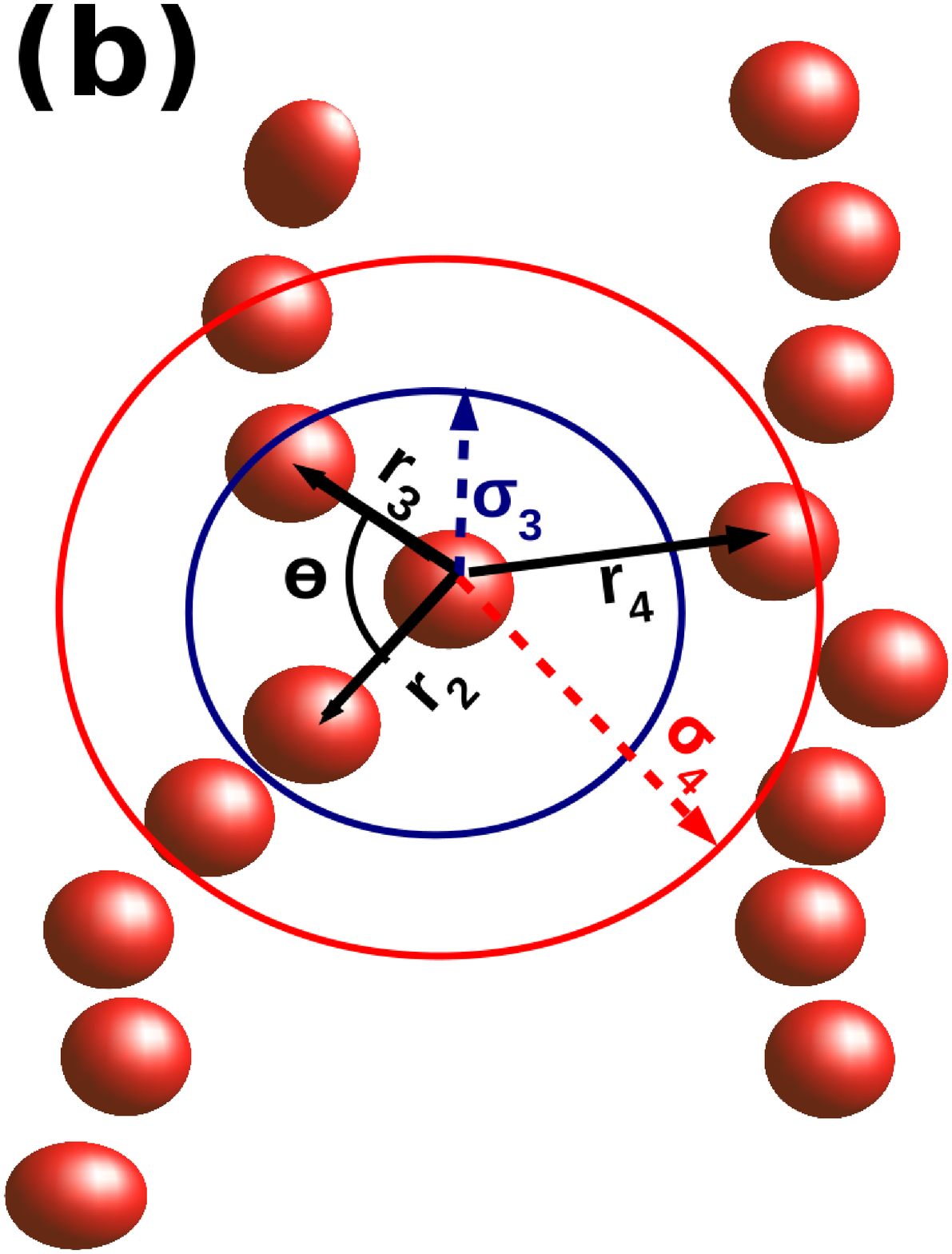}
\caption{ \label{fig1}  $(a)$ shows a plot of the two body  and three body potential between 2 monomers 
$V_2 (r_2) + V_3(r_2,r_3) $; $r_2$ ($r_3$) is the
 distance between monomer $1$ and $2$ ($1$ and $3$).  
Refer $(b)$ for a schematic diagram for $r_2,r_3,r_4, \theta$ and $\sigma_3, \sigma_4$.
The inset figure in $(a)$ shows the  repulsive potential $V_4$ on a fourth monomer for distance 
$r_4 < 2^{1/6} \sigma_4$ from central particle. Note $\sigma_4 > \sigma_3$.  
}
\end{figure}

\begin{figure*}
%\begin{figure}
\includegraphics[width=0.15\textwidth]{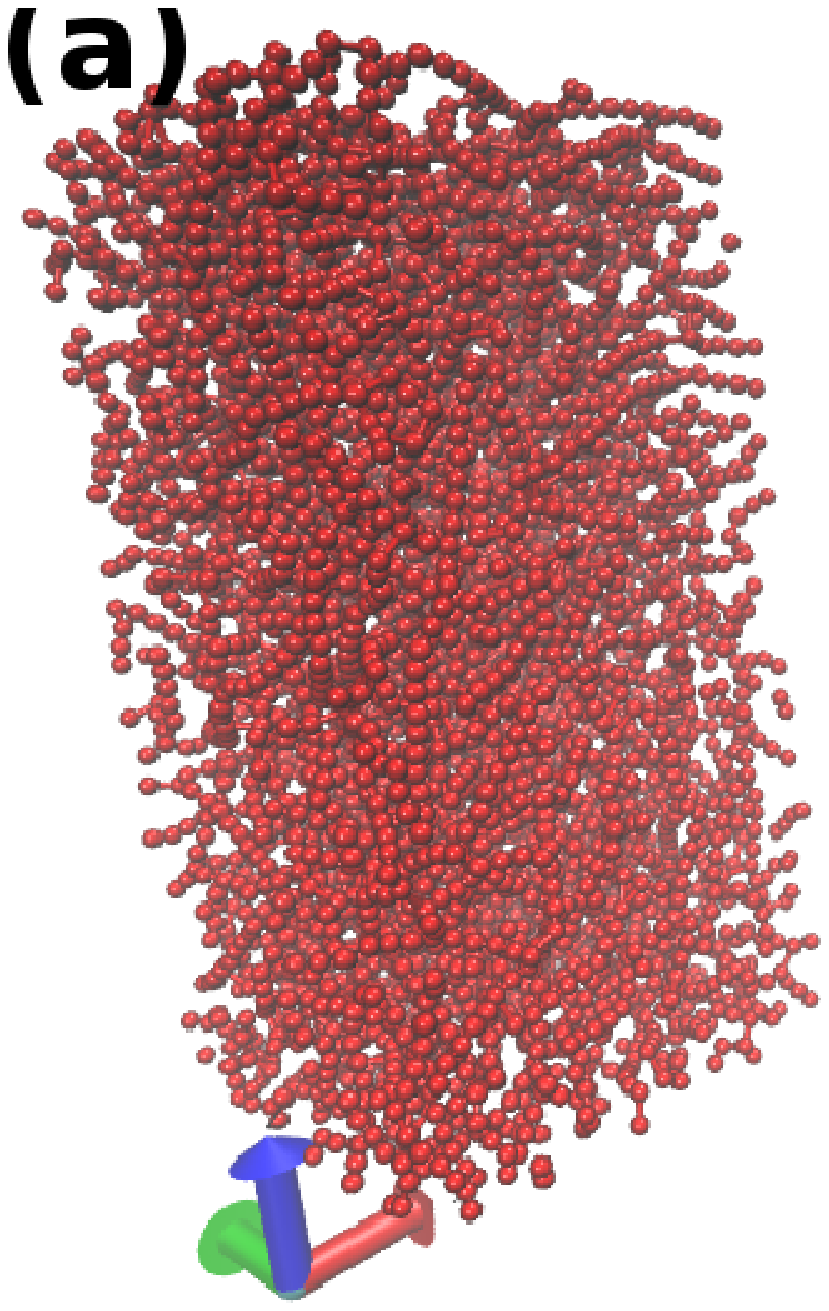}
\includegraphics[width=0.18\textwidth]{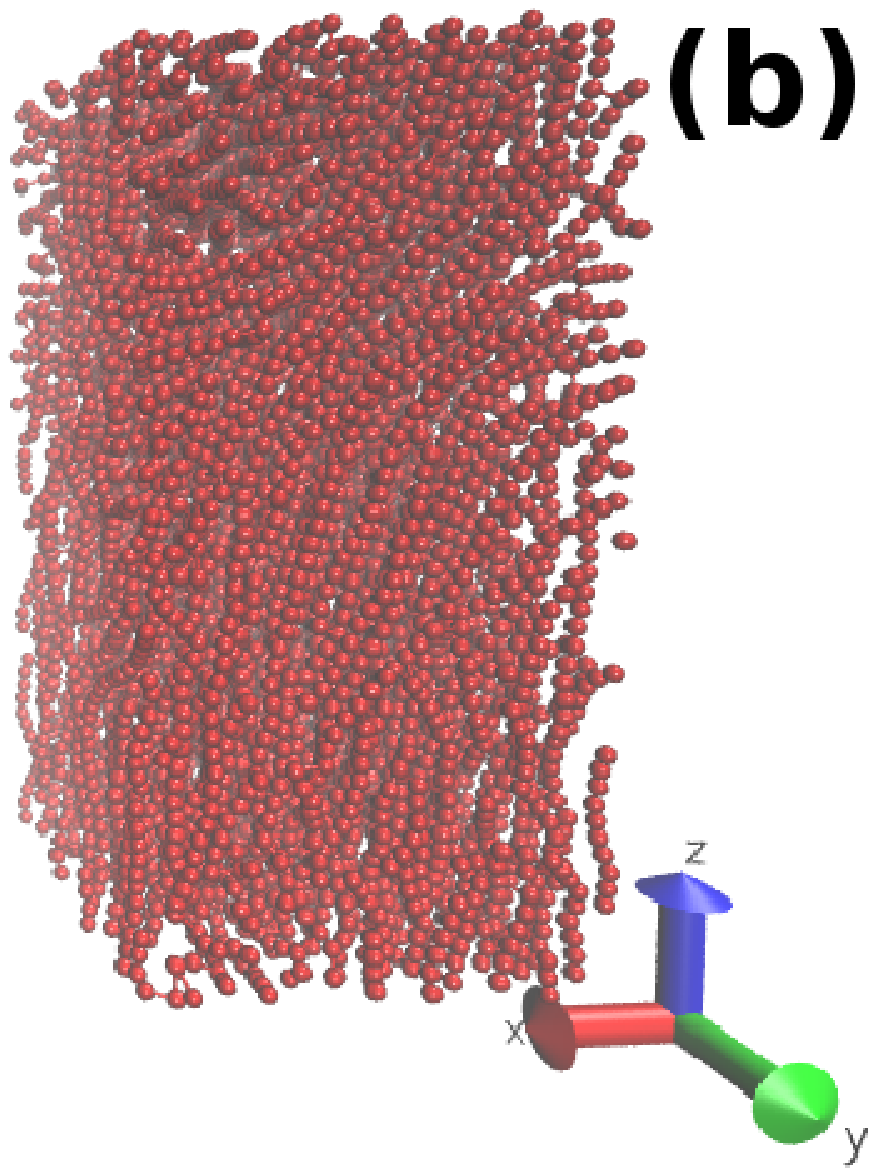}
\includegraphics[width=0.19\textwidth]{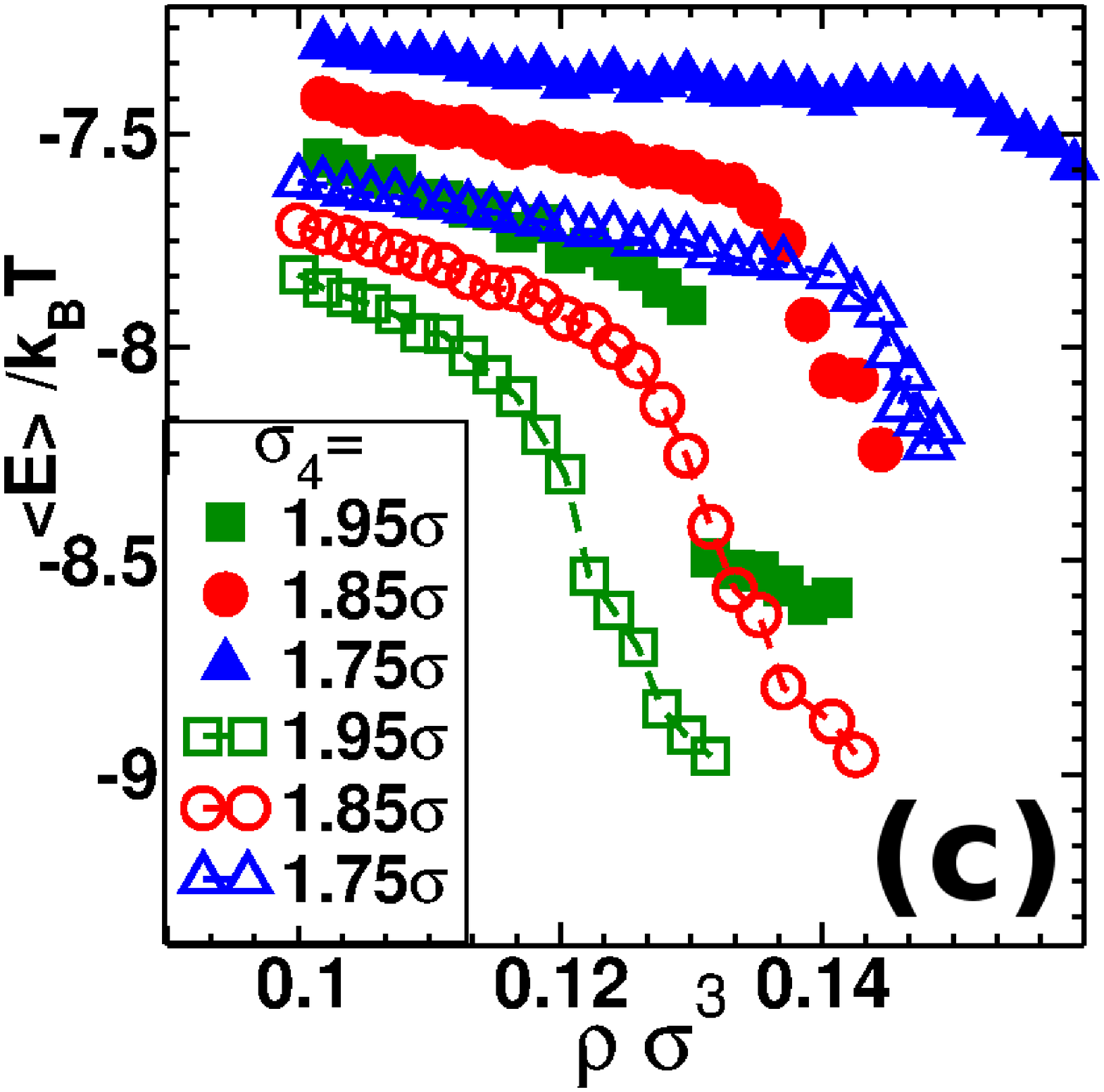}
\includegraphics[width=0.19\textwidth]{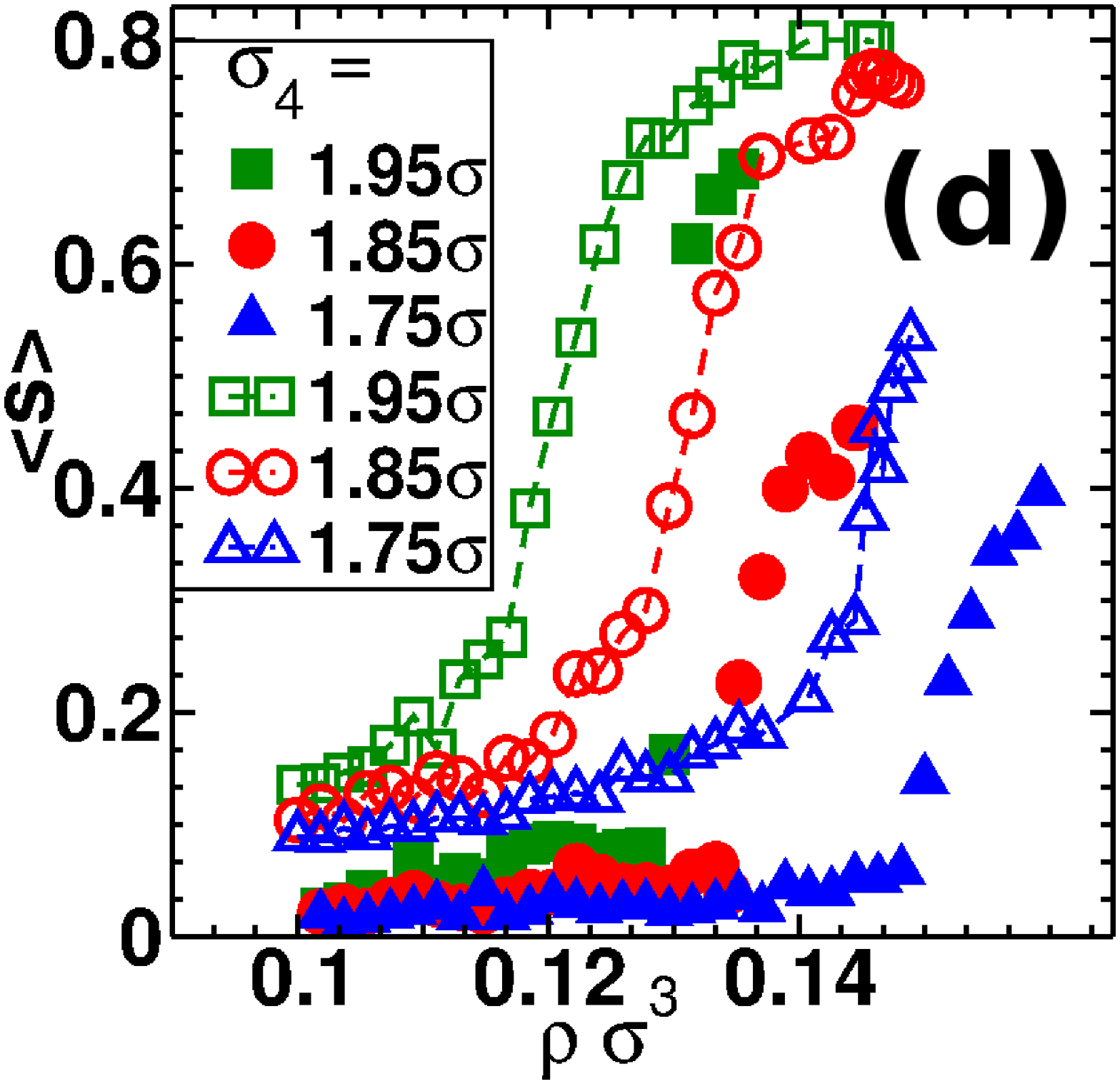}
\includegraphics[width=0.19\textwidth]{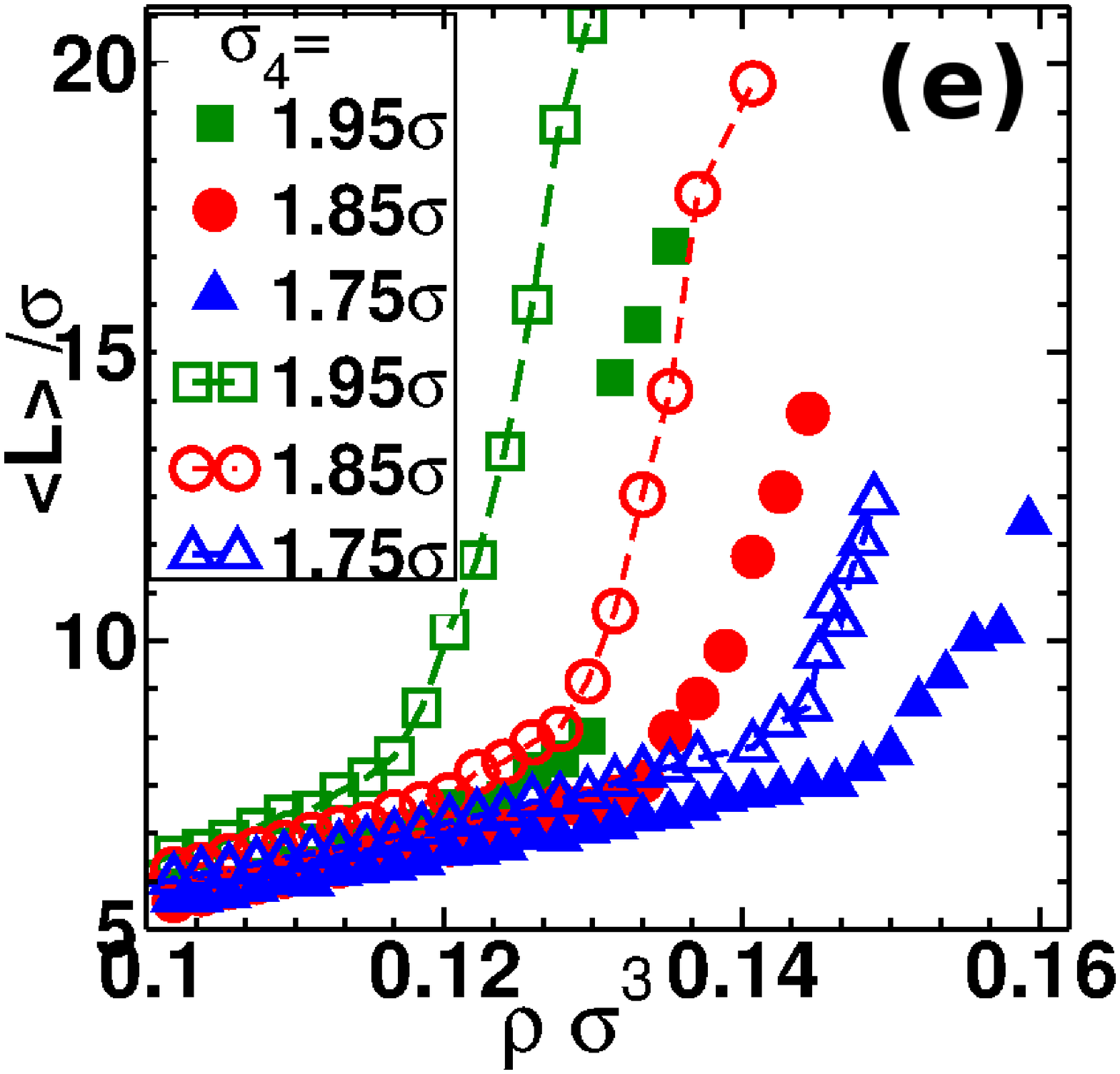}
\caption{ \label{fig2} Representative snapshots of $(a)$ isotropic and $(b)$ nematically ordered phase of  self-assembling 
monomers forming worm-like micelles(WM). The number of monomers are $6000$ and $7500$, respectively, in a $30 \times 30 \times 60 
\sigma^3$ box for a small symmetry breaking field $B_n^2 = 0.025 k_BT/\sigma^2$ and $\sigma_4 = 1.95 \sigma$. 
To identify the isotropic-nematic transition densities 
plots of average energy $\left< E \right>$,  average length $\left< L \right>$ and  the nematic order parameter $\left< s \right> $ 
versus the number density $\rho \sigma^{-3}$ is shown in subfigures $(c)$,$(d)$ and $(e)$, respectively.  
Data is shown for $B_n^2 = 0.025 k_B T/\sigma^2$ (filled symbols) and $0.1  k_B T/\sigma^2 $ (empty symbols).}
%\end{figure}
\end{figure*}

Here $r_2$ ($r_3$) is the distance between monomers $1 \& 2$ (monomers $1 \& 3$), 
the exponential term in Eqn.\ref{eq1} creates a maxima in the $V_2 (r)$ at $r \approx 1.75 \sigma$ 
(refer Fig.\ref{fig1}a); we set $r_c=2.5 \sigma$, $\sigma_3 = 1.5 \sigma$ and define $\sigma_3$ to be the cutoff 
distance above which the bond between monomers is considered broken.  The 3-body interaction 
$V_3 (r_{2},r_{3}) $, which models semiflexibility of chains, sets in only when a monomer has 
at least two bonded neighbours at distances $r_2,r_3 < \sigma_3$ (see Fig\ref{fig1}b). The angle
 $\theta$ is subtended between $\vec{r_2}$ and $\vec{r_3}$.  We set $\epsilon_1 = 1.34 \epsilon$, 
$\epsilon_3 = 6075 k_B T$ and $a = 1.72$.  A string of monomers can line up and form a chain and
the value of $\epsilon_3$ determines the measure of semi-flexibility of a chain.  The squared terms
which are a prefix to $sin^2(\theta)$ in Eqn \ref{eq2} ensure that the potential and the force go 
smoothly to zero as either of $r_{2}/\sigma_3 $ or $r_{3}/\sigma_3 \rightarrow 1$.  Note in 
Fig.\ref{fig1}a that the effective depth of the potential is a few times $k_B T$ and depends on 
an interplay between $V_2$ and $V_3$.

To prevent branching, we ensure that a fourth monomer does not approach a triplet of monomers and is 
repulsed by a additional potential $V_4(r_2,r_3,r_4)$, if the distance $r_4$ between the central monomer 
and the fourth monomer becomes less than $ \sigma_4$.  We choose $\sigma_4 = 1.75 \sigma$. 
\begin{equation}
V_4 = \epsilon_4 (1 - \frac{r_{2}}{\sigma_3})^2(1 - \frac{r_{3}}{\sigma_3})^2 \times V_{LJ}(\sigma_4,r_4) 
\label{eq3}
\end{equation}
$\forall r_4 < 2^{1/6} \sigma_4$,  and  $ \forall r_2,r_3 < \sigma_3$.  Here $V_{LJ}(\sigma_4,r_4)$ 
provides  purely repulsive interaction between monomers 1 and 4 by a suitably shifted and truncated
 Lennard Jones (LJ) potential.  The expression for $V_{LJ}$ is $(\sigma_4/r_4)^{12} -  
(\sigma_4/r_4)^6$.  The large value of $\epsilon_4=2.53 \times 10^5 k_BbT$  ensures sufficient 
repulsion even when both $ (1- r_{2}/\sigma_3)^2$ and $(1- r_{3}/\sigma_3)^2 \approx 0.01$; this  
corresponds to $r_{2}/\sigma_3=0.9$ and $r_{3}/\sigma_3=0.9$.  We explicitly checked that the 
chains do not have branches. The extra $V_4$ potential and the modified terms in Eqn. \ref{eq2} improves
upon the previous potentials used to model equilibrium polymers \cite{Chatterji2001,Chatterji2003,Prathyusha2013,Snigdha2010}.

\begin{figure*}
\includegraphics[width=0.2\textwidth,height=0.4\columnwidth,keepaspectratio=true]{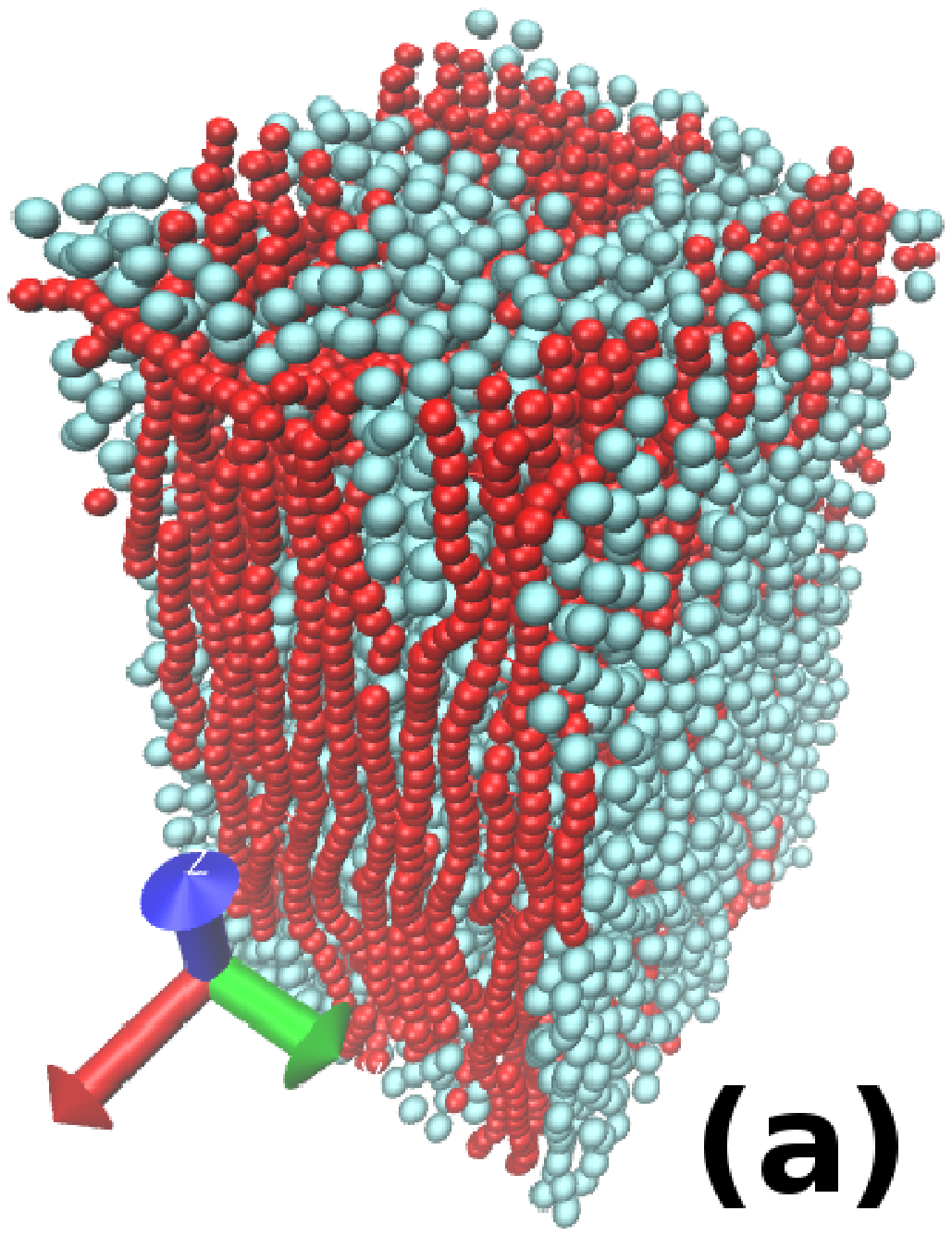}
\includegraphics[width=0.15\textwidth,height=0.4\columnwidth,keepaspectratio=true]{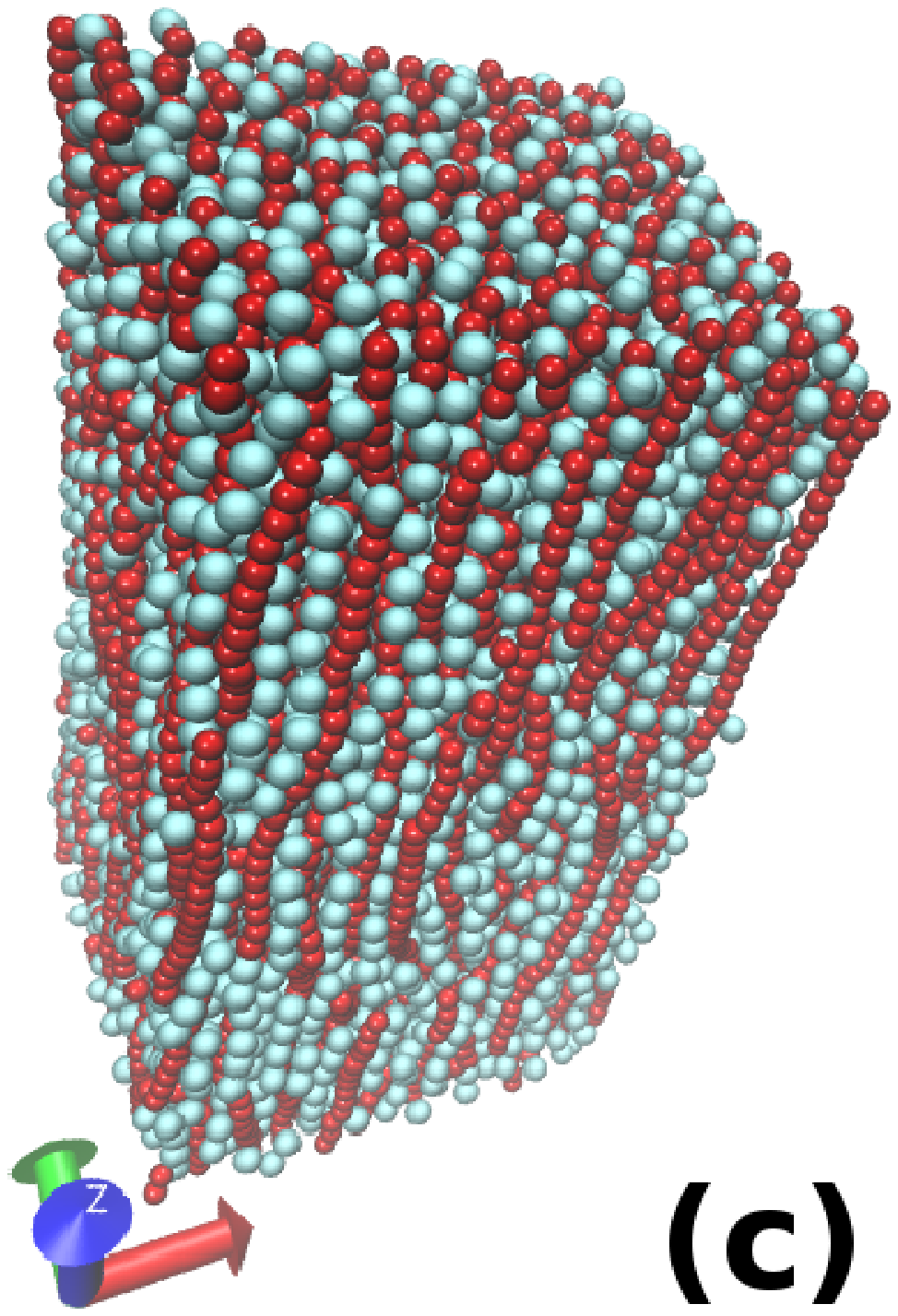}
\includegraphics[width=0.15\textwidth,height=0.43\columnwidth,keepaspectratio=true]{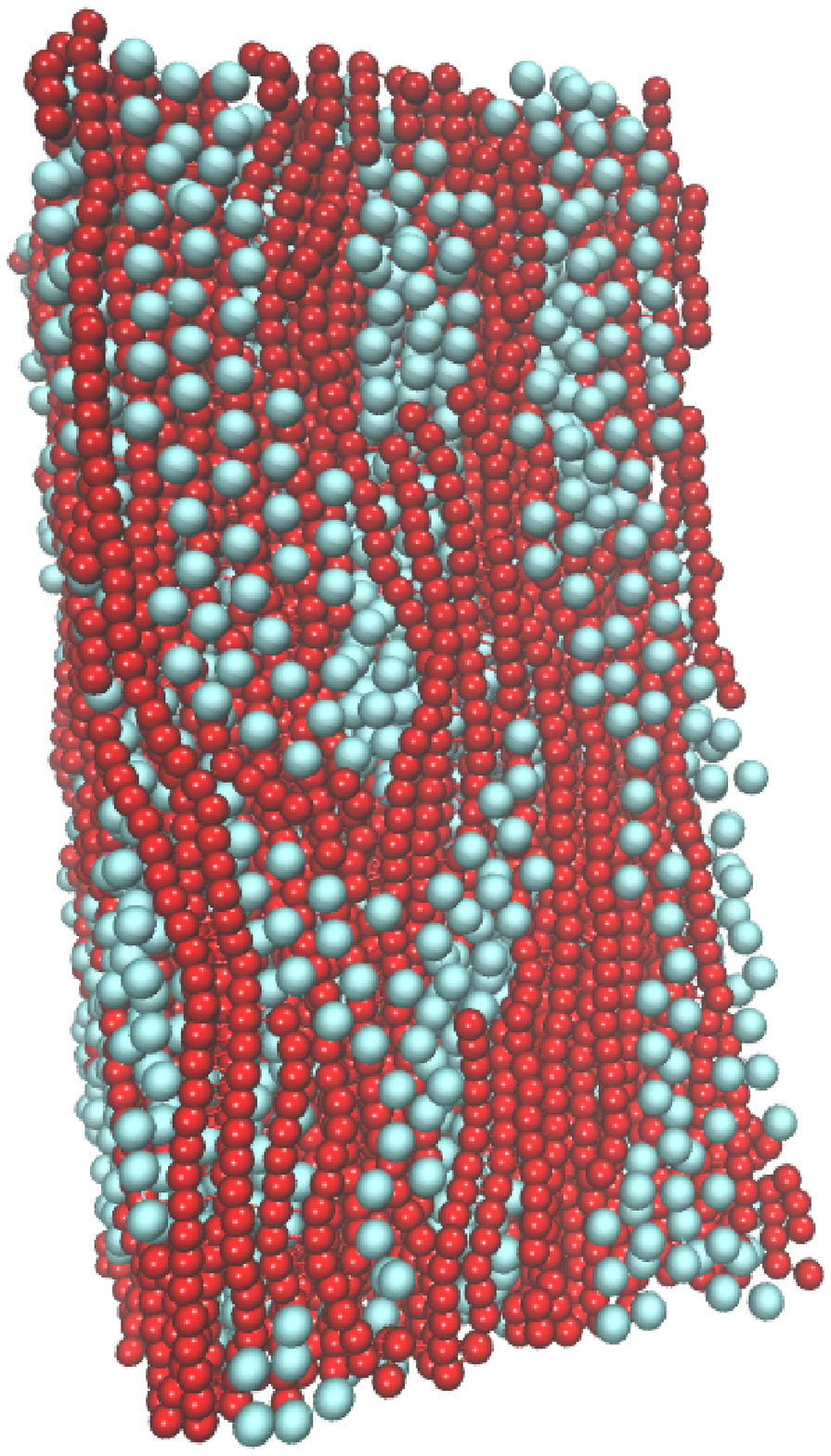}
\includegraphics[width=0.15\textwidth,height=0.4\columnwidth,keepaspectratio=true]{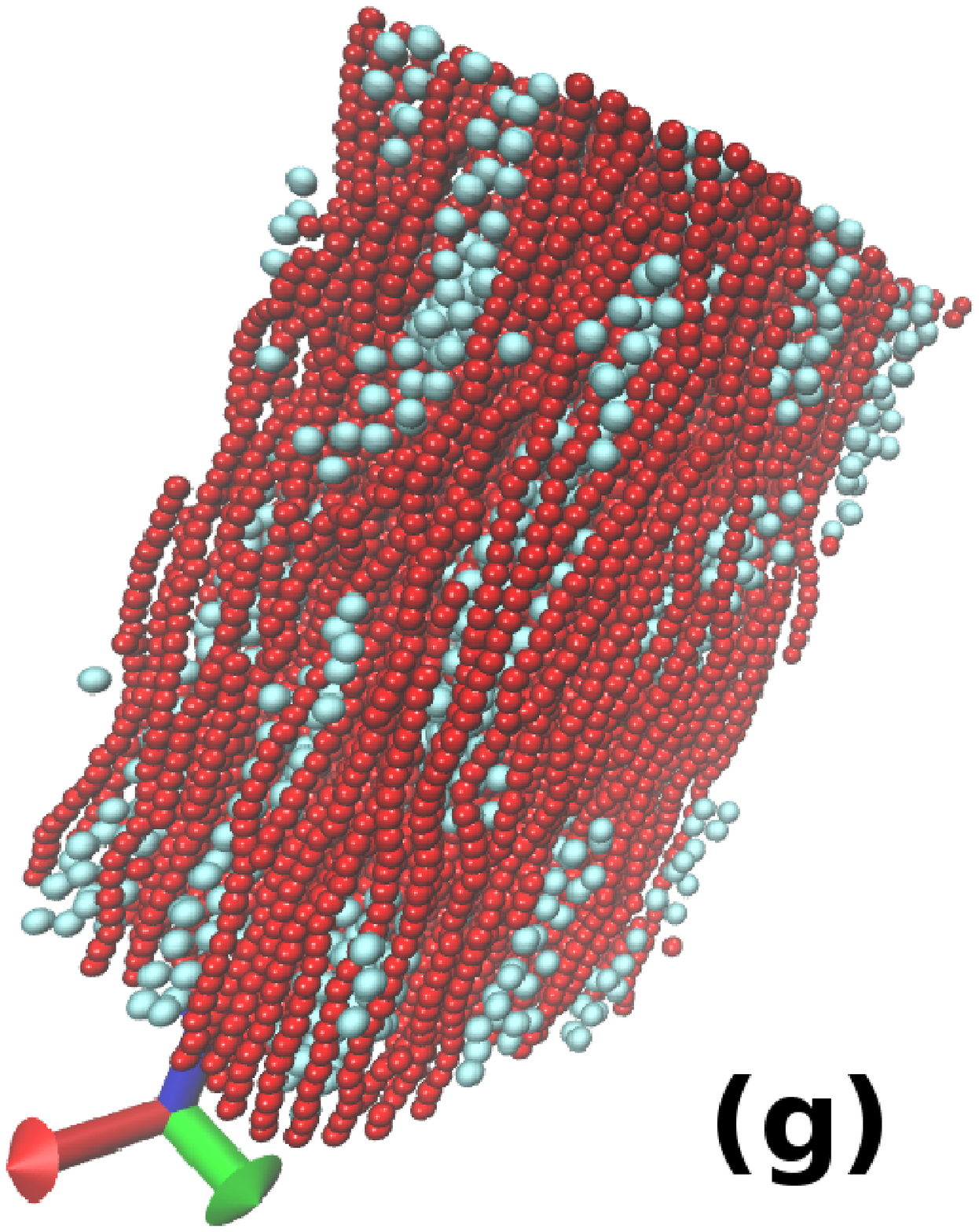}
\includegraphics[width=0.15\textwidth,height=0.4\columnwidth,keepaspectratio=true]{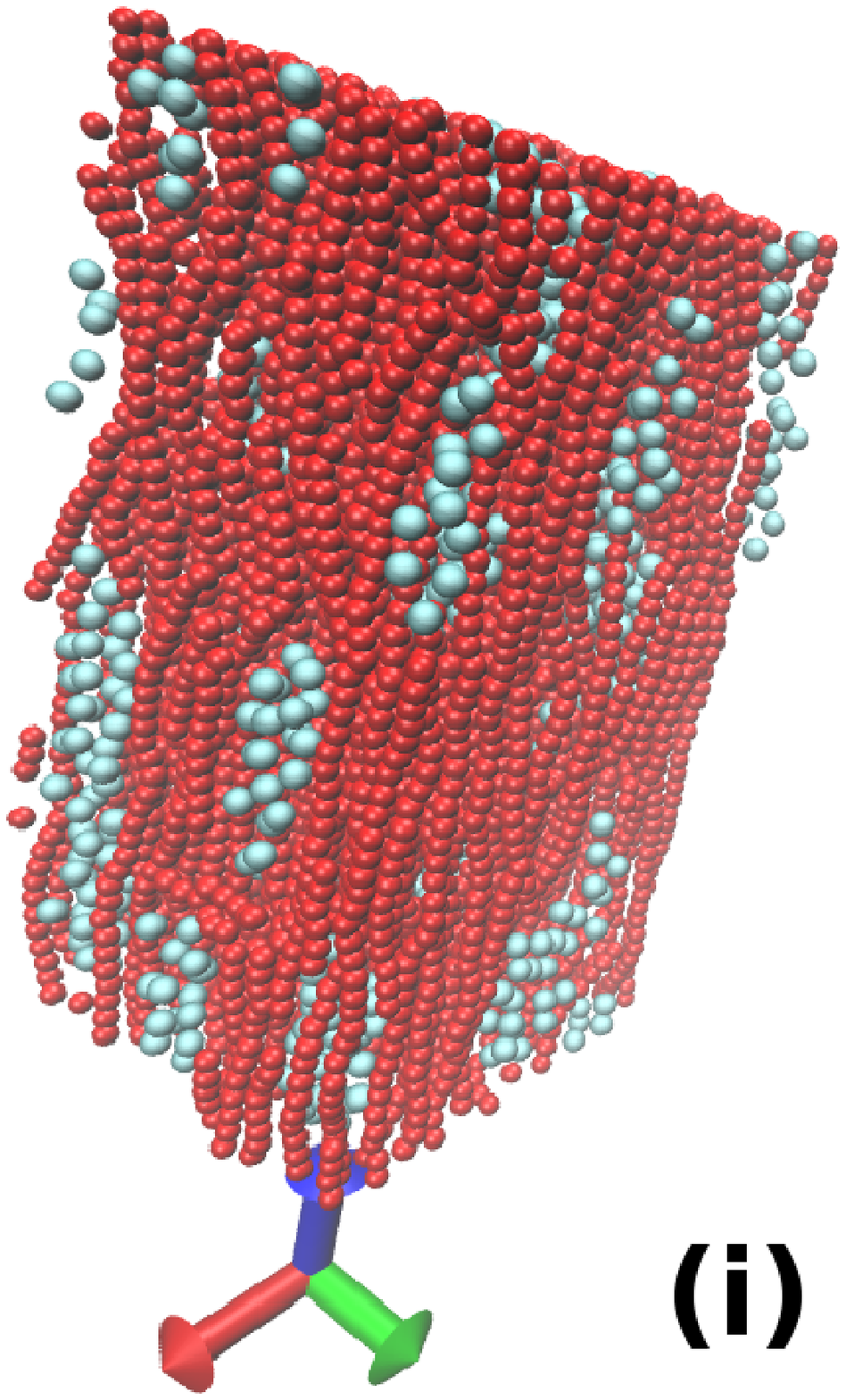}
\includegraphics[width=0.15\textwidth,height=0.4\columnwidth,keepaspectratio=true]{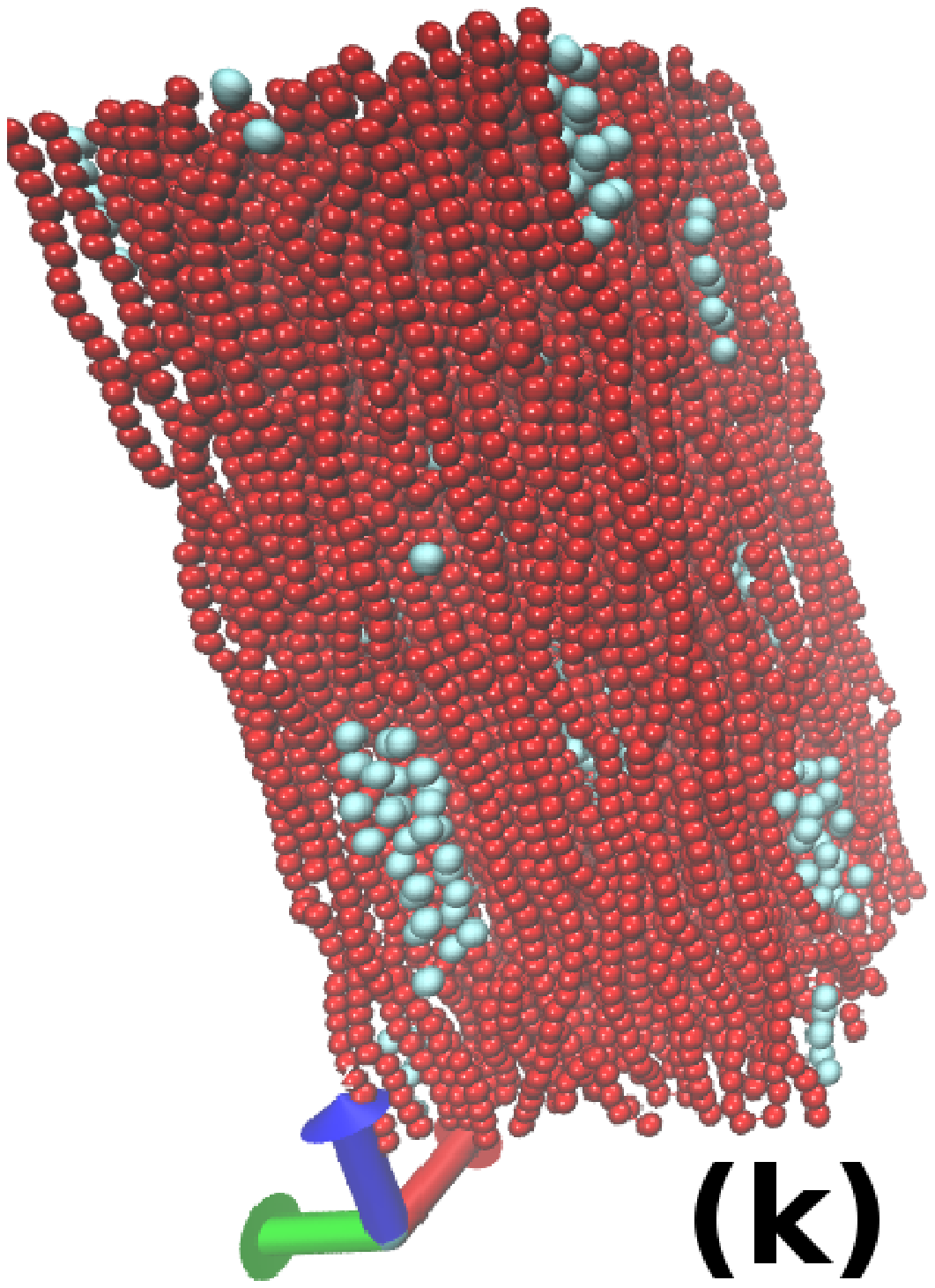}
\includegraphics[width=0.2\textwidth,height=0.4\columnwidth,keepaspectratio=true]{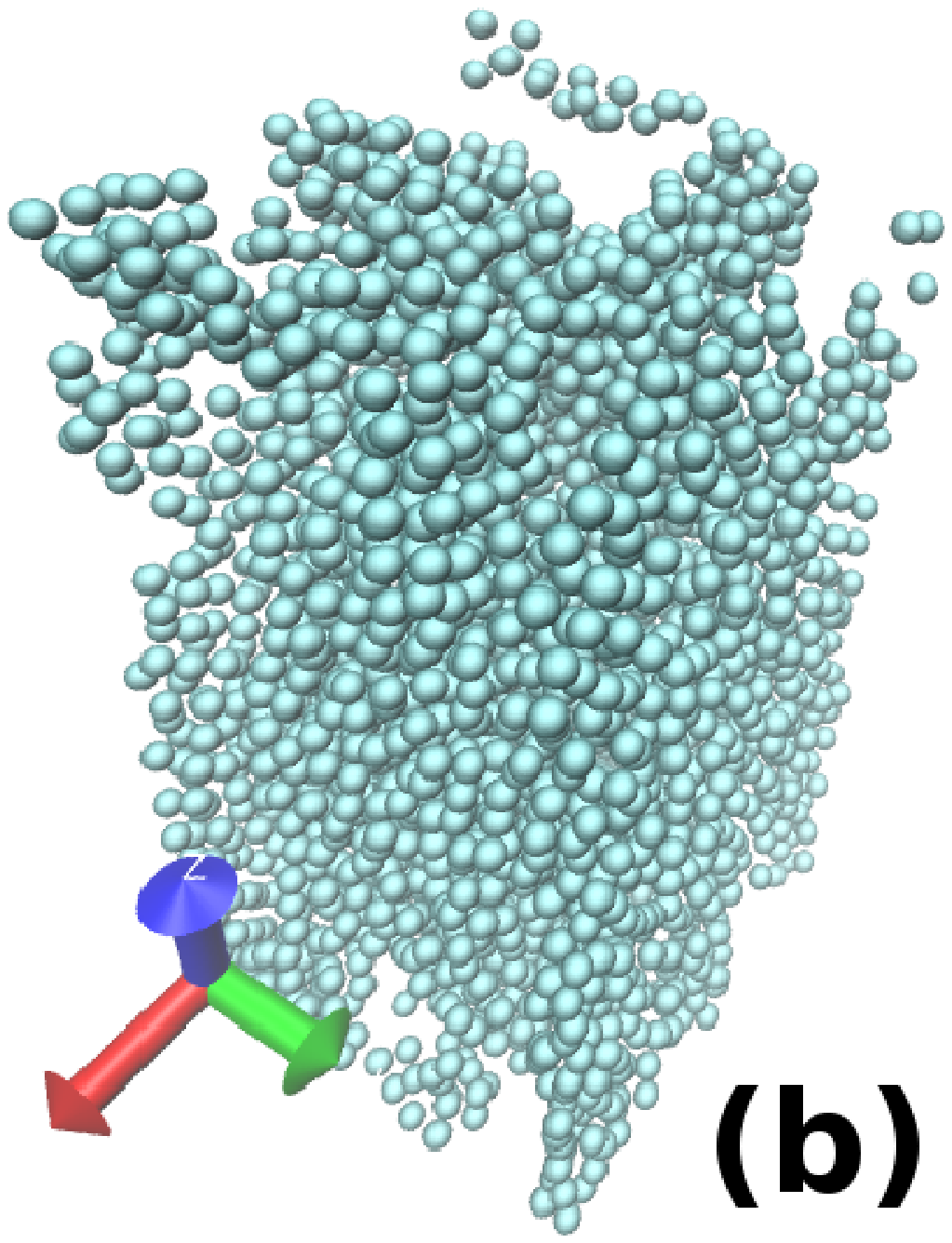}
\includegraphics[width=0.15\textwidth,height=0.4\columnwidth,keepaspectratio=true]{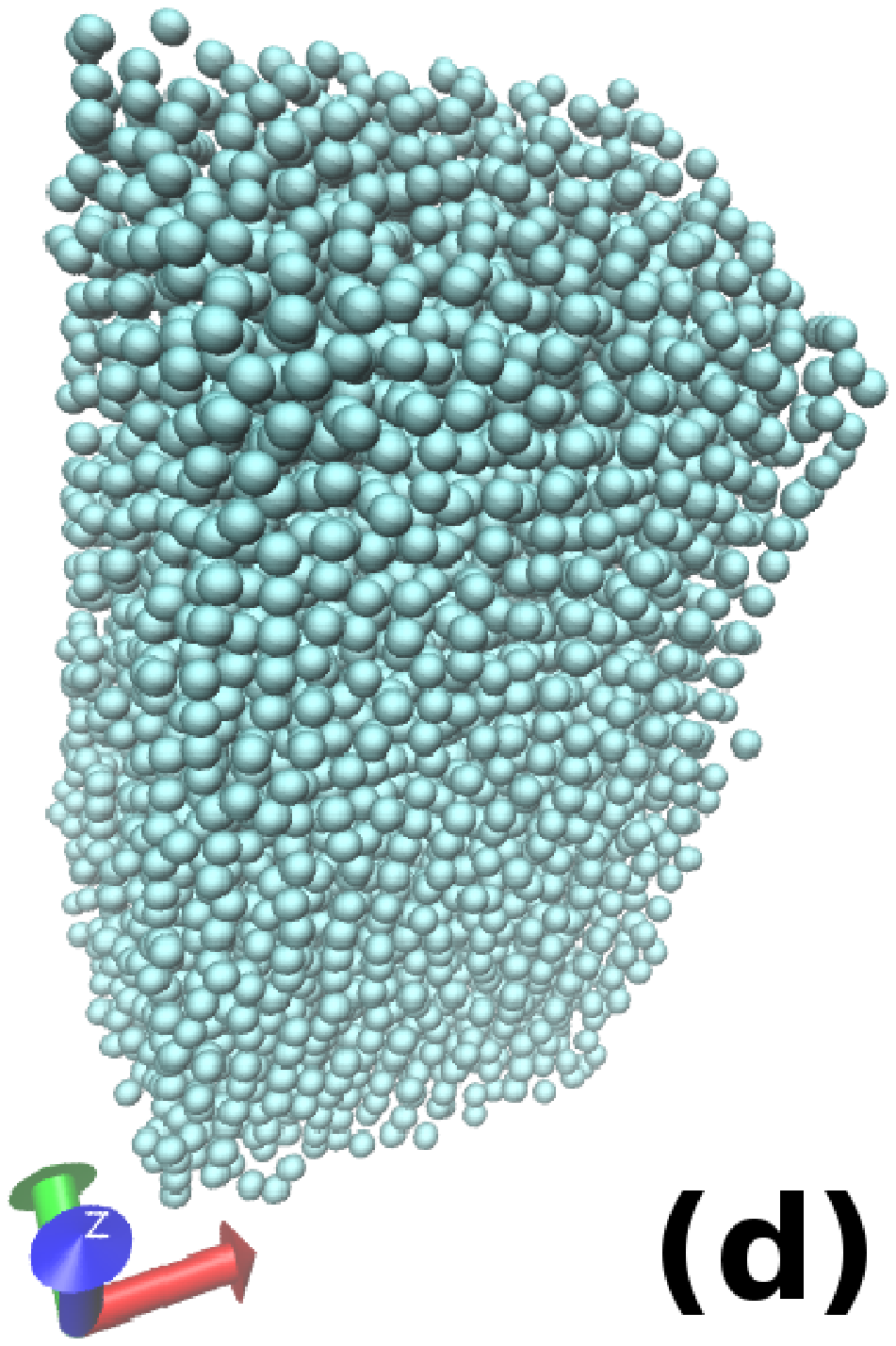}
\includegraphics[width=0.15\textwidth,height=0.43\columnwidth,keepaspectratio=true]{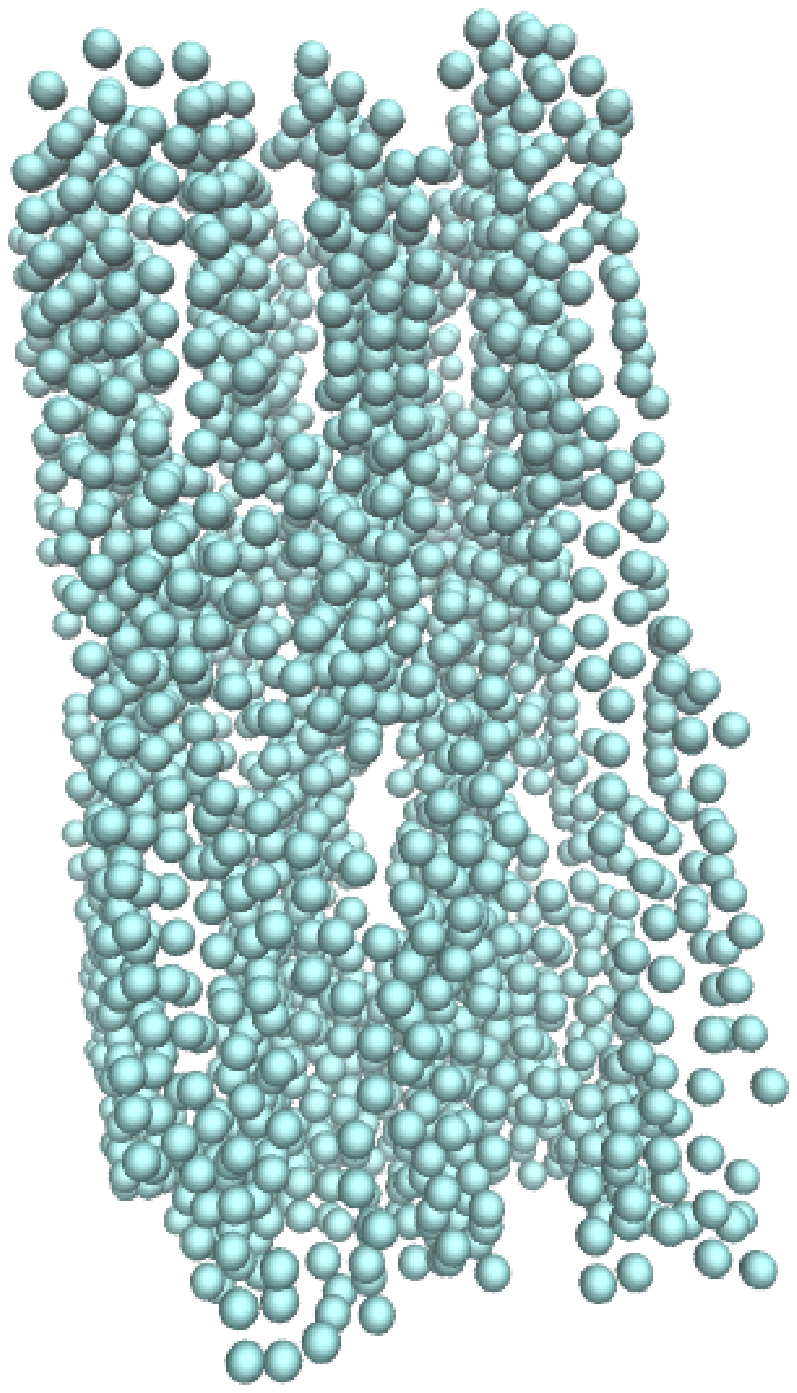}
\includegraphics[width=0.15\textwidth,height=0.4\columnwidth,keepaspectratio=true]{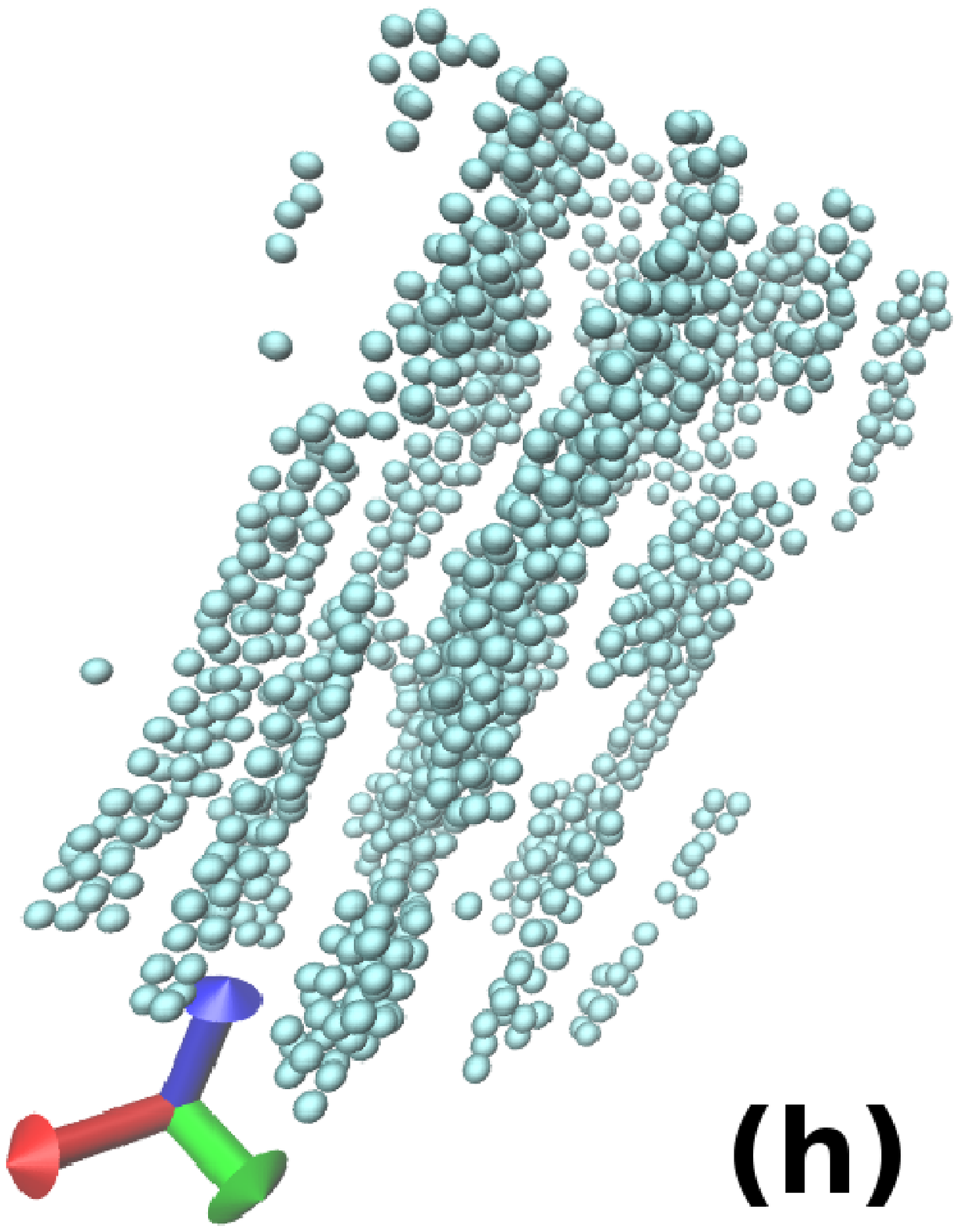}
\includegraphics[width=0.15\textwidth,height=0.4\columnwidth,keepaspectratio=true]{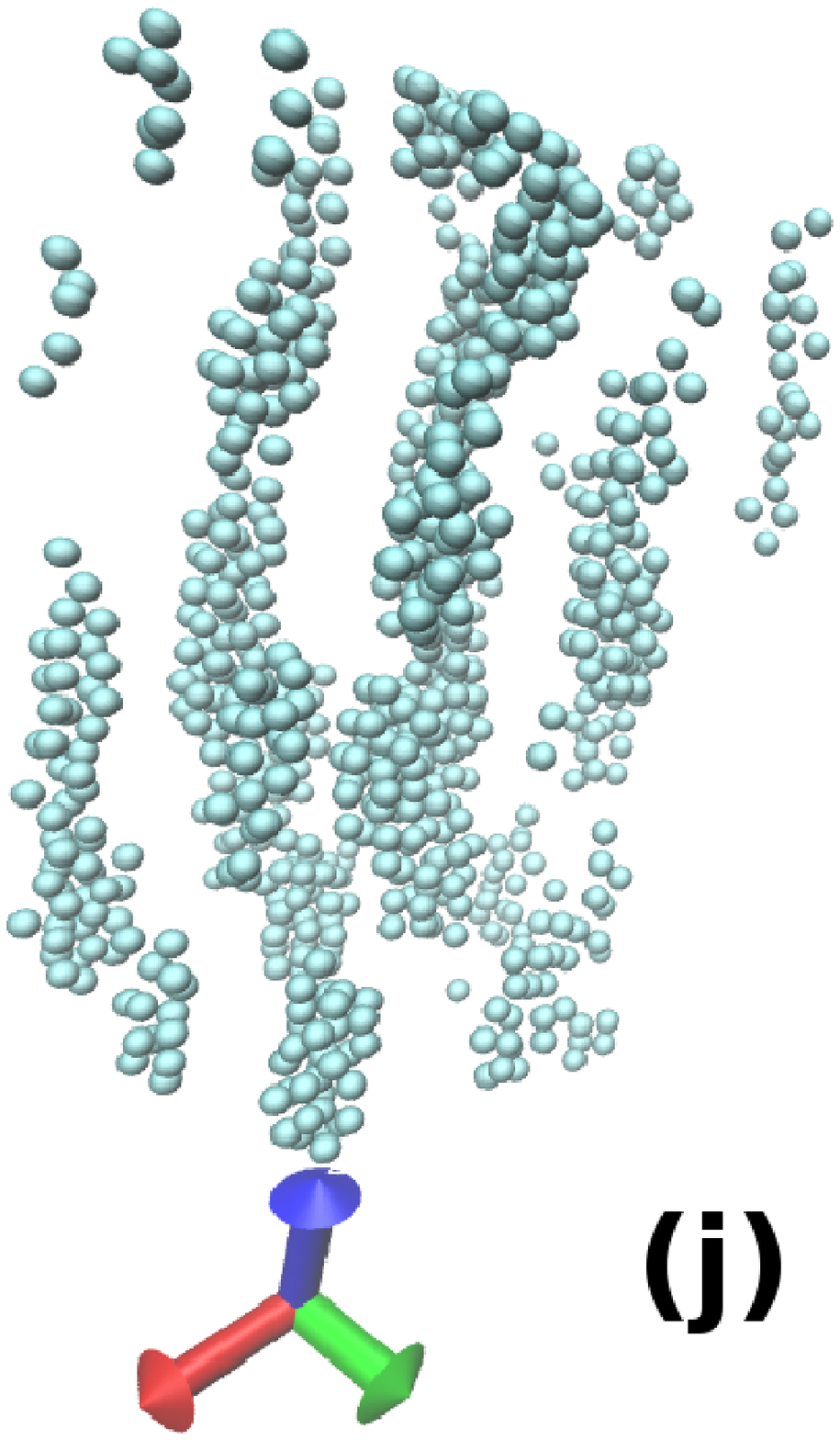}
\includegraphics[width=0.15\textwidth,height=0.4\columnwidth,keepaspectratio=true]{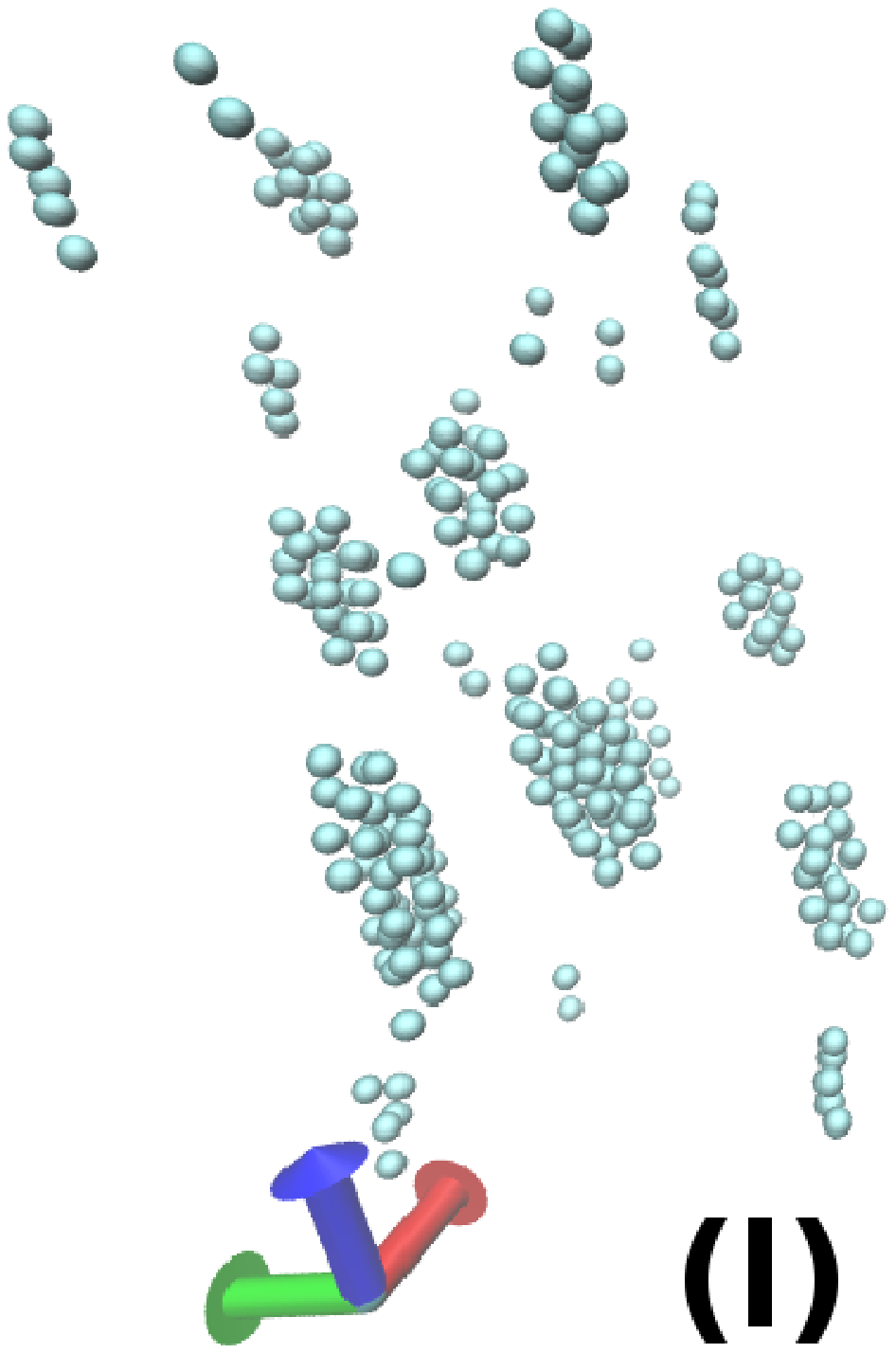}
\caption{ \label{fig3}
Snapshots of  WM chains (red) and structures formed by aggregates of nano particles-NP (blue) are shown. 
The snapshots in $b,d,f,h,j,l$ are exactly the same snapshots of $a,c,e,g,i,k$,  with monomers made invisible 
to have unhindered view of NPs.  $(a)$ and $(b)$ show an equilibrated phase separated configuration with 
4200 monomers and $\mu =-0.48 k_BT$ and $\epsilon_n = 10 k_BT$ 
for NPs with $\sigma_n =2 \sigma$ and $\sigma_{4n}=1.75$ in a $30 \times 30 \times 60 \sigma^3$ box.
Snaphots $(c)$ to $(l)$  show dynamically arrested configurations of monomers and NPs; for these 
the values of  $\mu, \epsilon_n$ and the NP radius  $\sigma_n$ is kept fixed at $16 k_BT, 14 k_bT$ and $2 \sigma$, 
respectively.  But $\sigma_{4n} = 1.5,2,2.5,2.75, 3.25 $ is varied for $(c),(e),(g),(i),(k)$.
The phases shown in snapshots $c,e,g,i,k$ are crystalline (Cr), percolating network (${\rm P_n}$),  elongated clusters (E), 
(E) but with shorter clusters, and aggregates (A), respectively. The volume fraction of NPs in $(a)$ is $0.21$ ($2712$ NPs),
for $c,e,g,i,k$ the NP-density is $0.38,0.17,0.092,0.061,0.018$, respectively.     
% $B_n^2 = 0.1 k_BT/\sigma^2$.
}
\end{figure*}

Figure \ref{fig2}$a$ and \ref{fig2}$b$ shows snapshots of $6000$ and $7500$ monomers, respectively, 
in a $30 \times 30  \times 60 \sigma^3$ simulation box after the system is equilibrated using Monte Carlo (MC)
Metropolis algorithm. The low density equilibrium configuration is a disordered isotropic phase with small 
chains. We have checked that an exponential distribution of chain lengths is obtained \cite{Cates1999}. 
To obtain the nematically ordered phase with long chains as shown in Fig. \ref{fig2}b, we have used a symmetry
breaking field $B_n$ which adds an energy $E_B = -(\hat{r}_{2} \cdot \vec{B}_n)^2$ to the Hamiltonian; $B_n$ acts only if 
$|\vec{r}_{2}| < \sigma_3$. We  use $B_n^2 = 0.025 k_BT/a^2$, which biases $\vec{r}_{2}$ to allign 
along $\hat{z}$.

In Figure \ref{fig2} $c,d$ and $e$ we have identified the range of densities over which the isotropic to 
nematic (I-N) transition occurs by plotting the average energy $\left< E \right> $ of the monomers, the nematic
order parameter $s = \left< 3 \cos^2(\phi) - 1/2 \right>$ and the average length of chains $\left< L \right>$ as a 
function of number density $\rho$ of monomers.  The angle $\phi$ is
the angle between a bond vector connecting adjacent monomers in a chain and  $\hat{z}$. 
All the quantities show a sharp increase/decrease  at the transition, i.e., for values 
of $\rho > 0.12 \sigma^{-3}$. An increase in $\sigma_4$, which  increases the volume excluded
 by the chains, makes the transition sharper and also shifts it to lower densities.
To assure ourselves of the robustness of our results, we carried out simulation for $B_n^2 = 0.1 k_BT/\sigma^2$ 
as well as lower value of  $B_n^2 = 0.025 k_BT/a^2$ in boxes of $30 \times 30 \times 30 \sigma^3$ and $30\times 30\times60 \sigma^3$, respectively.
A previous detailed study using a potential  similar to this model had 
established a weakly first order isotropic to nematic phase transition \cite{Chatterji2001}.  
 We can use the term $V_4$ to  vary EV of chains, {\rm and} avoid branching even for MD simulations.
We maintain $\rho= 0.126 a^{-3} $ ($6800$ monomers) at the I-N transition with $\sigma_4 = 1.75 \sigma$ in a  box 
of $30 \times 30  \times 60 \sigma^3$ and the $B_n^2 = 0.1  k_B T/\sigma^2$ fixed, for results presented hereafter.

\begin{figure}
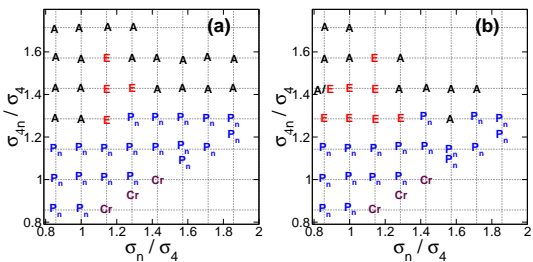

\includegraphics[width=0.4\columnwidth]{e14_mu16.eps}
\includegraphics[width=0.4\columnwidth]{e14_mu80.eps}
\caption{ \label{fig4}
The chart denoting the different dynamically arrested nano-structures formed as we change $\sigma_{4n}$ and $\sigma_n$ for 
$\epsilon_n =14 k_B T$ and $(a)$ $\mu = 16 k_b T$, ($b$) $\mu = 8 k_b T$. Refer Fig. \ref{fig3} for nano-structures nomenclature.
}
\end{figure}

 To study NP microstructure formation in a matrix of nematic polymers, we start introducing NPs amidst pre-formed 
nematically ordered chains. The monomers  self assemble into  ordered WMs within $10^5$ 
MC steps starting from a random initial configuration of 6800 monomers and 100 NPs. 
Then we attempt to add (remove) randomly chosen NPs $300$ times every $50$ MC steps for the next $19 \times 10^5$ MC steps. 
In one MC step, we attempt 
to change the position of all particles present by $\delta_i$, $\delta_i$ is a random number between $0$ and $0.25 \sigma$ 
and $i \in x,y,z$.  Nano-particles of diameter $\sigma_n$ interact with each other by the LJ potential suitably 
truncated at $2.5 \sigma_n$ and the interaction energy is $\epsilon_n$. 
% The interaction $V_{4n}$ between NP and micellar chains (with at least three monomers) is that  of EV interactions provided a repulsive 
%potential  similar to the expression of $V_4$ in Eqn. \ref{eq3}, but with $\epsilon_4, \sigma_4$
%and $r_4$ replaced by $\epsilon_{4n},\sigma_{4n}$ and $r_{n}$ for an approaching NP. 
The repulsive interaction between a NP and a monomer is given by the (suitably shifted) LJ-potential $V_{4n} = \epsilon_{4n} 
[(\sigma_{4n}/r_n)^{12} - (\sigma_{4n}/r_n)^{6}], \,\,  \forall r_n < 2^{1/6} \sigma_{4n} $, where $\epsilon_{4n}=30 k_BT$ and  $r_n$
is the NP-monomer distance.  We keep all the WM-chain parameters fixed and vary only the EV diameter $\sigma_{4n}$
between WM and NP, and $\sigma_n$.  

We perform grand canonical MC (GCMC) simulations  with the number of monomers fixed, 
but with an energy gain (loss) of $-\mu n_p (\mu n_p)$ for $n_p$ added (removed) 
NPs is the simulation box; $\mu$ is the chemical potential. 
The equilibrium phase of the micelles and NPs is the phase separated structure as seen in Fig. \ref{fig3}$a$ 
and $b$ for low densities of particles with  $\mu=-0.48k_B T$, $4800$ monomers, $\sigma_{4n} = 1.75 \sigma$, $\sigma_n=2 \sigma$
and $\epsilon_n=10 k_BT$.   By virtue of periodic boundary conditions, one can discern that there is 
only one single phase separated aggregate of monomeric chains.  For denser systems with $\mu > 0$, $6800$ 
monomers  and $\epsilon_n =14 k_BT$, we observe aggregation of NPs between WM-stacks. 
However,  kinetic barriers in these dense glassy systems are too high to enable the system to 
completely phase separate, 
but GCMC steps for NPs help overcome local energy barriers enabling them to form  aggregates: dynamically arrested
phase separating structures.  Within $10^6$ MC steps, the structure of NP-aggregates within the
micellar matrix gets nearly fixed, with slow changes in energy (less than $2.5 \%$) and addition of only 
$\sim 100$ NPs over the next $10^6$ MC steps \cite{supplement}. 

In Figure \ref{fig3} we show representative snapshots of self-assembled nano-structures formed by the aggregates 
of NPs in a matrix of nematically ordered WM chains. We keep $\epsilon_n=14 k_bT$, $\mu=16 k_B T$ and the NP radius 
$\sigma_n=2 \sigma$ of NPs fixed, and gradually increase the excluded volume parameter $\sigma_{4n}$ of the monomer-NP
interaction. At the lowest value of $\sigma_{4n}=1.5 \sigma$ (Fig.\ref{fig3}$c,d$), we get a phase where the NPs are 
arranged in a periodic manner separated by chains of WM monomers. 
The monomers and NP  form  crystalline domains (${\rm Cr}$) with alternate  positions of NP and monomer chains.
We calculated the structure factors for NPs which  confirm a crystalline structure (refer \cite{supplement}).  
This phase occurs whenever the condition $\sigma/2 + \sigma_n/2 = \sigma_{4n}$ is satisfied. {\em Moreover}, 
enough free volume should be available for the NPs to fill up all the possible lattice sites. 
The  free volume is dependent on both $\sigma_{4n}$ and $\sigma_n$ as well on the number density $\rho$ of monomers.
We have fixed  $\rho $ at $0.126 \sigma^{-3}$ and $\sigma_4$ at $1.75 \sigma$, 
but the number of NPs can adjust to fill up the  available space between WM chains. 
For other combinations of $\sigma_{4n}$ and $\sigma_n$, we get near perfect lattice arrangement of NPs; 
the reader can confirm this in  \cite{supplement}. 

If now $\sigma_{4n}$ is increased to $2 \sigma$ from $1.75 \sigma$,  a new structural arrangement of NPs is formed where the NPs start to
phase separate to form percolating clusters of NPs which span throughout the system: refer Fig.\ref{fig3}$e$ and $f$.
Kinetic trapping prevents complete phase separation, and we observe  NP clusters separated by stacks of WM chains lead to
morphologies  which is akin percolating network (${\rm P_n}$) of NPs.
If the WM phase was dissolved away at the end of the $P_n$ formation as in \cite{Hegmann2007,Wang2009}, 
 a porous scaffold of NP micro-structure would be retained similar to what is seen in Fig.\ref{fig3}f.
Increase of $\sigma_{4n}$ will lead to elongated structures conjoined at fewer points in space with fewer NPs in the system, and finally
at $\sigma_{4n}= 2.5 \sigma$, we have even fewer NPs which now  form non-percolating elongated
clusters (${\rm E}$) of NPs spanning the $\hat{z}$ direction: refer Fig.\ref{fig3}$g$ and $h$.
Clusters grow along the nematic direction to minimize elastic energy costs paid by nematically ordered WMs  to
accomodate NP clusters. A further increase in the value of $\sigma_{4n}$ to $2.75 \sigma$ leads to shorter and thinner
elongated structures  as shown in Fig.\ref{fig3}$i$ and $j$.
Finally, small aggregates (${\rm A}$)  dispersed in nematic matrix are found for
even larger values of $\sigma_{4n} = 3.25$ some of which are rod like, as seen in Fig. \ref{fig3}$k$ and $l$.
For further increase in $\sigma_{4n}$, it is not possible to introduce NPs in the nematic matrix. 
The decrease in number of NPs as we change $\sigma_{4n}$ is shown in \cite{supplement}.

The chart in Fig.\ref{fig4}$a$ and $b$ summarizes the dynamically arrested nano-structures that we get as we systematically 
vary $\sigma_{4n}$ and $\sigma_n$ for two values of $\mu$, viz, $\mu=16 k_BT$ and  $\mu=8 k_BT$.
The quantities $\sigma_{4n}$ and $\sigma_n$ have been normalized by $\sigma_4=1.75 \sigma$.
We have chosen $\sigma_{4n}$ (the EV separation between 
NP and micellar chain), to be different from $\sigma_4$ (the EV separation between two parallel chains of WM) to have 
an independent handle on changing NP-WM interactions keeping the volume fraction of micellar chains fixed.
In an experimental realization, $\sigma_{4n}$ could be different from 
that of $\sigma_4$ due to differences in microscopic interactions \cite{Hegmann2007}. We ensure that
$\sigma_{4n}$ is such that $\sigma_{4n} \ge (\sigma + \sigma_{n})/2$. As mentioned before, we get ${\rm Cr}$
arrangement of NPs when the condition $\sigma_{4n} = (\sigma + \sigma_{n})/2$ is satisfied. At times, the 
lattice arrangement does not span the system and forms crystalline domains instead. 
We get perfect crystal structures for $\epsilon_n = 30 k_BT$, see \cite{supplement}.
As $\sigma_{4n}$ is increased, we get the ${\rm P_n}$ phase and then the ${\rm A}$ phase for all values of $\sigma_n$,
but there also exists  an island of elongated clusters of NPs (${\rm E}$) in the phase diagram. 
For certain parameter values, these elongated clusters are perfectly rod like, refer \cite{supplement}.
%The elongated NP clusters we show in Fig.\ref{fig4}$(e)$ are rod-like 
%For $\mu = 8 k_BT$ corresponding to 
%$\sigma_n/\sigma_4 =1.14$ and $\sigma_{4n}/\sigma_4 =1.57$. 

In summary, we have used semi-grand-canonical simulations to demonstrate the aggregation and growth of nanoparticle
clusters with different morphologies. The NP clusters get dynamically arrested within  self-assembled chains of semiflexible worm-like
micelles. The different nanostructures obtained are (a) a crystalline arrangement of NPs and monomers,
(b) percolating networks of aggregated NPs creating a porous structure, 
(c) elongated rod like structures of NPs of variable length and aspect ratios, 
and (d) smaller clusters of different shapes and sizes. In contrast to systems theoretically investigated previously, 
our choice of the size and densities of NPs are such that NP clusters and micellar matrix mutually affect 
and modify each other's local morphology and  structure. 
We can get different nano-structures by varying $\mu$ of NPs, the EV of micellar chains seen by NPs as well 
as the NP radius. In future we plan to vary monomer density as well.
%We hope that our work motivates experimentalists to try this new methodology to explore the formation of nano-structures by forcing NPs to
%mix with self assembling nematogens with densities close to the isotropic-nematic transitions boundaries.     

AC acknowledges the use of the compute-cluster of Nano-Science unit in IISER,  funded
by DST, India by project no. SR/NM/NS-42/2009. AC thanks K. Guruswamy for
very useful discussions.

%\nocite{*}
%\bibliographystyle{plain}
%\bibliographystyle{apsrev4-1}
%\bibliographystyle{apsauth4-1}
%\bibliography{/home/apratim/Documents/mendeley_desktop/file_30_sep}

\end{document}